\documentclass[showpacs,prb,preprint]{revtex4}
\usepackage{amssymb}
\usepackage{graphicx}

\begin{document}

\title{Electron spin relaxation via flexural phonon modes in semiconducting
carbon nanotubes}
\author{K. M. Borysenko, Y. G. Semenov, and K. W. Kim}
\affiliation{Department of Electrical and Computer Engineering,
North Carolina State University, Raleigh, NC 27695-7911}

\author{J. M. Zavada}
\address{U.S. Army Research Office, Research Triangle Park, NC 27709}

\begin{abstract}
This work considers the g-tensor anisotropy induced by the flexural
thermal vibrations in one-dimensional structures and its role in
electron spin relaxation. In particular, the mechanism of
spin-lattice relaxation via flexural modes is studied theoretically
for localized and delocalized electronic states in semiconducting
carbon nanotubes in the presence of magnetic field. The calculation
of one-phonon spin-flip process predicts distinctive dependencies of
the relaxation rate on temperature, magnetic field and nanotube
diameter. Comparison with the spin relaxation caused by the
hyperfine interaction clearly suggests the relative efficiency of
the proposed mechanism at sufficiently high temperatures.
Specifically, the longitudinal spin relaxation time in the
semiconducting carbon nanotubes is estimated to be as short as 30
$\mu$s at room temperature.
\end{abstract}


\maketitle

\section{Introduction}

Unique electronic, mechanical and structure properties of\ carbon
nanotubes (CNTs) continue to be a focus of extensive investigation
to date (see Refs.~\onlinecite{Dresselhaus} and
\onlinecite{Ando05} as well as the references therein). For one,
it is considered to be the ultimate system for continued "scaling"
beyond the end of the semiconductor microelectronics
roadmap.\cite{Dekker} At the same time, recent
studies~\cite{Tsukagoshi99,Sahoo05} drew attention to other
important applications of CNTs.  Their crystalline properties with
low or no impurity incorporation allow the injection and use of
electrons with polarized spin as an added variable for
computation. Thus, CNTs are an ideal medium for the development of
the new emerging field of spintronics.~\cite{Meh,Yang}  Further,
the anticipated long spin relaxation times allows coherent
manipulation of electron spin states at an elevated temperature,
opening a significant opportunity for spin-based quantum
information processing.~\cite{LossDiVincenzo,KaneNature1998}
Clearly, spin dependent properties of CNTs warrant a comprehensive
investigation with the spin relaxation times/processes as one of
the most crucial.

Recently, electron spin relaxation in the CNTs was examined
theoretically by considering the hyperfine interaction (HFI) with
nuclear spins $I=1/2$ of $^{13}$C isotopes (with the natural
abundance of $1.10\%$).~\cite{SemenovHFI2007} The HFI is thought
to be the most important spin relaxation process with the role of
spin-orbit coupling routinely dismissed on the ground of its
weakness in the carbon-based structures. However, the calculation
result predicts the electron spin relation time of the order of
one second; a number much longer than that experienced
experimentally.

In the present work, we reexamine the role of spin-lattice
interaction on electron spin relaxation in semiconducting CNTs.
Although the $g$-tensor anisotropy caused by the spin-orbit
coupling does not lead to spin relaxation, it can provide an
effective mechanism when combined with specific thermal vibrations
of the CNT (namely, the flexural phonon modes). Indeed, any
bending of the CNT is accompanied by local rotation of the tube
axis and, thus, the principle axes of $g$-tensor. Consequently,
the electron Zeeman energy under a magnetic field will be affected
by thermal vibrations through the local angular displacement of
the tube axis, leading potentially to electron spin relaxation.

The rest of this paper is organized as follows. In Sec.~II, the
spin-phonon interaction Hamiltonian is derived in terms of
flexural vibrations and the CNT $g$-tensor components. In
Secs.~III and IV, this Hamiltonian is applied to the problem of
spin-flip relaxation in a magnetic field for electrons localized
in a finite segment of a CNT (analogous to a quantum dot) and for
delocalized electrons. Finally, Sec.~V discusses the numerical
results through comparison with the HFI-induced spin-flip
processes.  A range of parameter space is identified where the
proposed mechanism may provide a dominant contribution.

\section{Hamiltonian of the spin-phonon interaction}

Consider an electron under a magnetic field $ {\mathbf B}$.  The
Zeeman Hamiltonian  in an undisturbed CNT can be written in the
frame of reference with the axis $z$ directed along the tube as
\begin{equation}
H_{Z}^{(0)}=g_{\shortparallel }\mu _{B}B_{z}S_{z}+g_{\perp }\mu
_{B}(B_{x}S_{x}+B_{y}S_{y}),  \label{eq1}
\end{equation}%
where $\mu _{B}$ is the Bohr magneton, $\mathbf{S}$ an electron
spin,  $g_{\perp }=g_{xx}=g_{yy}$ and $g_{\shortparallel }=g_{zz}$
with the principal values of electron $g$-tensor $ g_{xx}$,
$g_{yy}$, $g_{zz}$.  In the presence of flexural thermal vibrations,
the system loses the axial symmetry. If the curvature induced by
flexural deformation is small (compared to the CNT diameter), one
can introduce local Cartesian axes $x^{\prime }$, $y^{\prime }$,
$z^{\prime }$ at each point along the CNT (Fig. 1). The angle
between the $z$ and $z^{\prime }$ axes is defined as $\theta =\theta
(z)$. Then, the Zeeman Hamiltonian in the presence of flexural
vibrations can be expressed in the local coordinate system (denoted
as $ H_{Z}^{\prime }$) by simple substitution $x$, $y$, $ z$
$\rightarrow$ $ x^{\prime }$, $y^{\prime }$, $z^{\prime }$ in
Eq.~(\ref{eq1}) while retaining the same parameters
$g_{\shortparallel }$ and $g_{\perp }$. When one transforms $
H_{Z}^{\prime }$ back to the laboratory coordinate $x $, $y$, $z$,
the corresponding Hamiltonian $H_{Z}$ shows explicit dependence on
the angular displacement. To proceed further we fix the angle
$\varphi $ between $x$-axis and $z-z^{\prime }$ plane and take into
account the smallness of $\theta $. Keeping the linear terms on
$\sin \theta $ , the Hamiltonian can be reduced
to the form $H_{Z}=H_{Z}^{(0)}+H_{s-ph}$ with perturbation%
\begin{equation}
H_{s-ph}=\Delta g\mu _{B}[(B_{x}S_{z}+B_{z}S_{x})\cos \varphi
+(B_{y}S_{z}+B_{z}S_{y})\sin \varphi ]\sin \theta ,  \label{eq2}
\end{equation}%
where $\Delta g=g_{\shortparallel }-g_{\perp }$.

Equation (\ref{eq2}) shows that fluctuations of $z^{\prime }$ axis leads to
fluctuating effective field that can mediate a spin relaxation. The angular
displacements $\theta $ and $\varphi $ are immediate from the transversal
deformation $\mathbf{u}$, so that

\begin{equation}
\sin \theta \cos \varphi =\frac{du_{x}}{dz};\hspace{0.02in}\sin \theta \sin
\varphi =\frac{du_{y}}{dz}.  \label{eq3}
\end{equation}%
In turn, the $\mathbf{u=u(}z)$ can be expressed in terms of flexural modes.
The latter are the two transversal acoustic modes that have orthogonal
polarizations $\mathbf{e}_{1}\perp $ $\mathbf{e}_{2}$ (we fix their
direction along $x$- and $y$- axes correspondingly) and possess a quadratic
dispersion for mode frequency
\begin{equation}
\omega _{\varkappa ,q}=\beta q^{2}  \label{eq3a}
\end{equation}%
with wave number $q$ and polarization $\varkappa =x,y$.\cite{Mahan2002}
Parameter $\beta $ in Eq. (\ref{eq3a}) is defined by the diameter $d_{t}$ of
CNT and its elastic properties,%
\begin{equation}
\beta =\lambda d_{t}v_{t},  \label{eq3b}
\end{equation}%
where $\lambda =0.56$ (Ref.\onlinecite{Mahan2002}) and $v_{t}=1.4\hspace{%
0.02in}10^{6}$ cm/s is a transverse sound velocity.\cite{Oshima88} The
quadratic dispersion law (\ref{eq3a}) results in singularity in phonon low
energy density of states that impacts the dependencies on magnetic field and
temperature as it will be shown below.

If we represent the $\mathbf{u(}z)$ in second quantization form \cite{Kittel}
and substitute it to the Eq. (\ref{eq3}) and then to Eq. (\ref{eq2}), after
some algebra one can find%
\begin{equation}
H_{s-ph}=\Delta g\mu _{B}\sum_{\varkappa =x,y}\sum_{q}iqe^{iqz}\sqrt{\frac{%
\hbar }{2\varrho _{1}A_{0}\omega _{\varkappa ,q}}}(B_{z}S_{\varkappa
}+B_{\varkappa }S_{z})(a_{\varkappa ,q}+a_{\varkappa ,-q}^{\dagger }),
\label{eq4}
\end{equation}%
where $\varrho _{1}$ is a linear density of CNT with length $A_{0}$, $%
a_{\varkappa ,q}^{\dagger }$ and $a_{\varkappa ,q}$ the operators of
creation and annihilation of the phonon with wave number $q$ and
polarization $\varkappa $.

Hereinafter we use only matrix elements of spin operators that describe the
spin-flip transitions between the eigenstates $\left\vert +\right\rangle $
and $\left\vert -\right\rangle $ of Hamiltonian $H_{Z}^{(0)}$(\ref{eq1}).
The diagonal matrix elements can also be responsible for the phase spin
relaxation providing the finite phonon lifetime,\cite{SK07} we however skip
this possibility assuming that phase spin relaxation is conditioned by
spin-flip processes.

The symmetry imposes the independence of the problem on azimuth direction
angle of a magnetic field. For definiteness sake we fix the $\mathbf{B}$
-direction so that $B_{x}=\sin \alpha $, $B_{y}=0$ and $B_{z}=\cos \alpha $,
where $\alpha $ is a vectoral angle between $\mathbf{B}$ and the tube axis $%
z $ (Fig. 1). Thus diagonalization of $H_{Z}^{(0)}$ gives rise to
the $\left\vert +\right\rangle =\left( \cos \frac{\alpha }{2},\sin
\frac{\alpha }{2}\right) $
and $\left\vert -\right\rangle =\left( -\sin \frac{\alpha }{2},\cos \frac{%
\alpha }{2}\right) $. Doing so we ignore a small difference in $%
g_{\shortparallel }$ and $g_{\perp }$. As a result on can find $\left\langle
-\left\vert B_{z}S_{x}+B_{x}S_{z}\right\vert +\right\rangle =\frac{1}{2}%
B\cos 2\alpha $ and $\left\langle -\left\vert
B_{z}S_{y}+B_{y}S_{z}\right\vert +\right\rangle =\frac{i}{2}B\cos \alpha $.

The Eq. (\ref{eq4}) is a basic equation that describes the electron spin
interaction with flexural modes in CNT. It will be applied for analysis of
spin-lattice relaxation in the cases of localized and delocalized electrons
in next two sections.

\section{Spin relaxation of localized electrons}

Consider a localized electron in the bias potential applied perpendicular to
the semiconductor CNT. Parabolic shape for such potential is a reasonable
approximation if we analyze only the ground electronic state.\cite%
{LossDiVincenzo,KaneNature1998} That assumes quite low temperatures, $%
k_{B}T\ll \hbar \omega _{0}$, compared with energy space $\hbar \omega _{0}$
to the first excited state. Under these approximations the envelope wave
function reads\cite{Ando05} $\Psi _{0}=\chi \psi _{T}(\xi )\psi _{L}(\eta )$%
, where $\xi $ and $\eta $ are the curvilinear coordinates associated with
the circumference and longitudinal length of the deflection curve of CNT, $%
\chi $ is a two-fold amount for $\psi $-function amplitudes at the A and B
atoms of a primitive cell. We consider small bends of CNT and join $\eta $
with $z$ axis so that longitudinal part of the $\Psi _{0}$ takes the form%
\cite{Sapmaz06}
\begin{equation}
\psi _{L}(z)=\sqrt{\frac{2}{\pi d_{0}}}e^{-z^{2}/d_{0}^{2}},  \label{eq01}
\end{equation}%
where $2d_{0}=2\sqrt{2\hbar /m^{\ast }\omega _{0}}$ is an extension of the
electron localization, $m^{\ast }=2\hbar ^{2}/3d_{t}\gamma $ is an effective
mass of the semiconducting CNT, $\gamma $ is a transfer matrix element. The
implicit forms of the $\chi $ and $\psi _{T}(\xi )$ are irrelevant to the
problem under consideration. Besides, we ignore the modifications of the
electronic states by the external magnetic field assuming that relevant
parameter\cite{Ando05} $(d_{t}/2a_{H})^{2}$ is rather small ($a_{H}=\sqrt{%
c\hbar /eB}$ is the magnetic length).

The averaging of electron-phonon interaction (\ref{eq4}) over the $\Psi _{0}$
can be reduced to the calculation of the form-factor $\Phi
(q/q_{0})=\left\langle \psi _{L}(z)|e^{iqz}|\psi _{L}(z)\right\rangle $, where $%
q_{0}\equiv \sqrt{8}/d_{0}$ and%
\begin{equation}
\Phi \left( x\right) =e^{-x^{2}}.  \label{eq02}
\end{equation}%
Thus, the Hamiltonian of spin-phonon interaction $V_{s-ph}=\left\langle \Psi
_{0}|V|\Psi _{0}\right\rangle $ for localized electron can be presented in
canonical form%
\begin{equation}
H_{s-ph}=\sum_{\varkappa =x,y}\sum_{q}V^{\varkappa ,q}(a_{\varkappa
,q}+a_{\varkappa ,-q}^{\dagger }),  \label{eq03}
\end{equation}%
where spin-depended operator is
\begin{equation}
V^{\varkappa ,q}=\Delta g\mu _{B}iq\sqrt{\frac{\hbar }{2\varrho
_{1}A_{0}\omega _{\varkappa ,q}}}\Phi \left( q/q_{0}\right)
(B_{z}S_{\varkappa }+B_{\varkappa }S_{z}).  \label{eq04}
\end{equation}

We are looking for the rate of longitudinal spin relaxation $%
T_{1}^{-1}=w_{+-}+w_{-+}$, which can be evaluated in terms of
probability of spin transition $w_{+-}$ ($w_{-+}$) from the spin
states $\left\vert +\right\rangle $ ($\left\vert -\right\rangle $)
to the state $\left\vert
-\right\rangle $ ($\left\vert +\right\rangle $). In turn, the $w_{+-}$ and $%
w_{-+}$ can be found by straightforward application of the Fermi
golden rule to the Eq. (\ref{eq03}). The final result of such
calculations takes the form
\begin{equation}
T_{1}^{-1}=\frac{2\pi }{\hbar ^{2}}\sum_{\varkappa
=x,y}\sum_{q}(2n_{\varkappa ,q}+1)\left\vert V_{+-}^{\varkappa
,q}\right\vert ^{2}\delta (\omega _{Z}-\omega _{\varkappa ,q}),
\label{eq05}
\end{equation}%
where $n_{\varkappa ,q}=\left\langle a_{\varkappa ,q}^{\dagger
}a_{\varkappa ,q}\right\rangle =1/\left[ \exp (\omega _{\varkappa
,q})-1\right] $ is the phonon population factor and $\hbar \omega
_{Z}=g\mu _{B}B$ is a Zeeman splitting, a weak anisotropy of
g-factor $g$ can be neglected here. The evaluation of the sum in Eq.
(\ref{eq05}) by means of Eq. (\ref{eq3a}) with
matrix elements of operators in Eq. (\ref{eq04}) leads the final result%
\begin{equation}
T_{1}^{-1}=\left( \frac{\Delta g}{g}\right) ^{2}\frac{\hbar \omega _{Z}^{3/2}%
}{16\varrho _{1}\beta ^{3/2}}\;F(\alpha )\Phi \left( \sqrt{\frac{\omega
_{Z}d_{0}^{2}}{4\beta }}\right) \coth \left( \frac{\hbar \omega _{Z}}{2k_{B}T%
}\right) ,  \label{eq06}
\end{equation}%
where $g\cong g_{\shortparallel },g_{\perp }$, the factor $F(\alpha )$ takes
into account the angular dependence of longitudinal relaxation,%
\begin{equation}
F(\alpha )=\frac{1}{2}\left( \cos ^{2}2\alpha +\cos ^{2}\alpha \right) .
\label{eq07}
\end{equation}

Equation~(\ref{eq07}) describes a prominent angular dependence of
spin-relaxation rate in CNT. It reaches the maximal value $F(\alpha
)=1$ at $\alpha =0$, i.e. a magnetic field applies parallel to the
axis of CNT, and reduces this value up to $F(\alpha )=0.22$ at
$\alpha =52.5^{\circ }$ (Fig. 2). In the case of perpendicular
orientation, $\alpha =90^{\circ }$, only modes with $x$ polarization
contribute to relaxation, i.e. $F(\alpha )=0.5$.

\section{Spin relaxation of delocalized electrons}

Let us consider the spin-flip relaxation for delocalized electrons in
semiconductor CNT. This case is most relevant to the spintronic
applications, so we extend the consideration up to room temperatures.
Assuming small displacements $\mathbf{u}$ we describe the longitudinal
component of orbital electronic states in the vicinity of the $K$ valley in
terms of plane waves with wave number $k$ counted off from the $K$ point of
the Brillouin zone, $\psi _{L}(z)=e^{ikz}/\sqrt{A_{0}}$. This approximation
immediately results in momentum conservation when interaction with phonons
describes the Hamiltonian (\ref{eq4}), i.e. $\left\langle k^{\prime
}|e^{iqz}|k\right\rangle =(2\pi /A_{0})\delta (q-k^{\prime }+k)$ and%
\begin{equation}
\left\langle k^{\prime }|H_{s-ph}|k\right\rangle =\sum_{\varkappa
=x,y}V_{k^{\prime } k}^{\varkappa }(a_{\varkappa ,-\Delta
k}+a_{\varkappa ,\Delta k}^{\dagger }),  \label{eq8}
\end{equation}%
where $\Delta k=k^{\prime }-k$, and
\begin{equation}
V_{k^{\prime } k}^{\varkappa }=i\text{sign}(\Delta k)\frac{\Delta g\mu _{B}}{%
2\pi }\sqrt{\frac{\hbar }{2\varrho _{1}A_{0}\beta }}(B_{z}S_{\varkappa
}+B_{\varkappa }S_{z}).  \label{eq8a}
\end{equation}%
The electron scattering in the vicinity of the $K^{\prime }$ valley gives
rise to the same results while intervalley scattering is negligible because
it entails the excitation of phonons with giant $q$.

The probability $w_{+-}^{k k^{\prime }}$ of spin-flip $\left\vert
+\right\rangle \rightarrow \left\vert -\right\rangle $ due to
electron scattering $k\rightarrow k^{\prime }$ on the phonons can
be found in terms of Fermi golden rule. The averaging over the
phonon thermal distribution results in
\begin{equation}
w_{+-}^{k k^{\prime }}=C\left[ n_{\Delta k}\delta \left( \Delta
E-\hbar \omega _{Z}-\hbar \beta \Delta k^{2}\right) +\left(
n_{\Delta k}+1\right)
\delta \left( \Delta E-\hbar \omega _{Z}+\hbar \beta \Delta k^{2}\right) %
\right]   \label{eq9}
\end{equation}%
where $C=\pi F(\alpha )(\Delta g\mu _{B}B)^{2}/2\varrho _{1}A_{0}\beta $, $%
\Delta E=\hbar ^{2}(k^{\prime }{}^{2}-k^{2})/2m^{\ast }$.
Similarly one can obtain the probability of spin flop $w_{-+}^{k
k^{\prime }}$. The net result for electron spin relaxation
$\overline{w}_{+-}$ assumes the averaging of the $w_{+-}^{k
k^{\prime }}$ over the electron initial state
occupation numbers $f_{k}^{+}$ and summation over non-occupied final states $%
k^{\prime }\left\vert -\right\rangle $,%
\begin{equation}
\overline{w}_{+-}=\sum_{k,k^{\prime }}f_{k}^{+}(1-f_{k^{\prime
}}^{-})w_{-+}^{k k^{\prime }}.  \label{eq10}
\end{equation}%
The summation over $k$, $k^{\prime }$ has to be replaced by
integration in
the common fashion. In the case of non-degenerate electrons,%
\begin{equation}
f_{k}^{+}=\frac{1}{A_{0}}\sqrt{\frac{2\pi \hbar ^{2}}{m^{\ast }k_{B}T}}\exp
\left( -\frac{\hbar ^{2}k^{2}}{2m^{\ast }}\right) .  \label{eq11}
\end{equation}%
One can see that the Eqs (\ref{eq9}) - (\ref{eq11}) are conveniently
expressed in terms of new variables $u=k+k^{\prime }$ and $v=k-k^{\prime }$
so that arguments of $\delta $-functions in Eq. (\ref{eq9}) transform to the
linear form on $u$. The integration over $u$ in Eq. (\ref{eq10}) becomes
trivial and gives rise to final expression for the spin relaxation rate $%
T_{1}^{-1}=\overline{w}_{+,-}+\overline{w}_{-,+}$ in the form%
\begin{equation}
T_{1}^{-1}=\frac{F(\alpha )}{\sqrt{8\pi }}\left( \frac{\Delta g}{g}\right)
^{2}\frac{\sqrt{m^{\ast }(k_{B}T)^{3}}}{\hbar \varrho _{1}\beta }h^{2}\cosh
\frac{h}{2}\cosh \frac{\varepsilon h}{2}I(h),  \label{eq12}
\end{equation}%
where we introduce the dimensionless parameters $h=\hbar \omega _{Z}/k_{B}T$
and $\varepsilon =2m^{\ast }\beta /\hbar $, and an integral function%
\begin{equation}
I(h)=\int_{0}^{\infty }\exp \left( -\frac{1+\varepsilon ^{2}}{4\varepsilon }%
x-h^{2}\frac{\varepsilon }{4x}\right) \frac{dx}{x\sinh (x/2)}.  \label{eq13}
\end{equation}%
Taking into account  Eq.~(\ref{eq3b}) and definition of $m^{\ast
}$, parameter $\varepsilon $ can be estimated as a number
$\varepsilon =0.02$ independent on $d_{t}$.

The integrant in Eq.~(\ref{eq13}) appears as a result of the
change of variable~$v$ by $x=\hbar \beta v^{2}/k_{B}T$. Asymptotic
behavior of the Eq. (\ref{eq13}) is $I(h\rightarrow 0)\rightarrow
8/\varepsilon h^{2}$, i.e. the integral diverges at zero magnetic
field ($h\rightarrow 0$). Actually this singularity has not
physical meaning. If electronic scattering is taking into account,
the wave numbers $k$ and $k^{\prime }$ are not longer accurate
quantum numbers and $\delta $-functions in the Eq. (\ref{eq9})
transform to delta-like functions with finite width and maximum.
This limits the value of integrant in Eq. (\ref{eq13}) and
function $I(h)$ at small $x$
and $h$ correspondingly. Similar situation was discussed in Refs \cite%
{Argyres,SemenovHFI2007}. Practically, the singularity treatment means that
once Zeeman energy $\hbar \omega _{Z}$ becomes of the order of electron
energy broadening $\Gamma $, the integral will reach a saturation, i.e. if
\begin{equation}
\hbar \omega _{Z}\lesssim \Gamma ,  \label{eq14}
\end{equation}%
then $I\left( \hbar \omega _{Z}/k_{B}T\right) \rightarrow I\left(
\Gamma /k_{B}T\right) $. As a result, zero magnetic field quenches
spin relaxation mechanism under consideration, $T_{1}^{-1}(\omega
_{Z}\rightarrow 0)\sim \left( \hbar \omega _{Z}\right) ^{2}I\left(
\frac{\Gamma }{k_{B}T}\right) \rightarrow 0$. We will not consider
this effect in more detail. Fig. 3 presents the function $h^{2}I(h)$
calculated with Eq. (\ref{eq13}) without considering the effect of
finite $\Gamma $. Note that the $h^{2}I(h)$ exponentially decreases
in the limit of large $h$.

\section{Numerical evaluation and discussion}

Equations~(\ref{eq06}) and (\ref{eq12}) define the spin relaxation
rate via flexural phonon modes for localized and delocalized
electrons. Alternative
approach based on the electron spin scattering on disordered nuclei spins $I=%
\frac{1}{2}$ of $^{13}C$ isotopes due to hyperfine interaction was
done in the Ref.\onlinecite{SemenovHFI2007}. Comparing these
mechanisms is performed for chirality vector (8,4) (respectively CNT
diameter and effective mass are $d_{t}=0.839$ nm and $m^{\ast
}=0.186\hspace{0.02in}m_{0}$, $m_{0}$ the free electron
mass)\cite{SemenovHFI2007} with following parameters: $\Delta
g=0.005
$, $g=2.015$, \cite{g-factor} $\varrho _{1}=\pi d_{t}\varrho _{2}$ where $%
\varrho _{2}=9.66\times 10^{-8}$ g/cm$^{2}$ is the mass of graphene sheet
per unit area.\cite{Ando05} Hereinafter all calculations are performed at $%
\mathbf{B}$ directed along the tube axis, i.e. $\alpha =0$.

Figure~4 illustrates the temperature and magnetic field dependencies
of spin relaxation rate of the electron localized at $2d_{0}=30$ nm.
The temperature dependence is typical for one-phonon process of
spin-lattice relaxation in
semiconductor quantum dot.\cite{SK07} At relatively high temperatures ($%
k_{B}T>\hbar \omega _{Z}$) the population factor for resonant
phonons \ describes the direct process $T_{1}^{-1} \sim T$ while the
temperature independent phonon radiation dominates at $k_{B}T<\hbar
\omega _{Z}$. Magnetic field dependence runs from zero through the
maximum, which is caused by form-factor at relatively short
wavelength of resonant phonon. Dependence on CNT diameter (Fig. 5)
stems from the explicit dependencies of form-factor, the $\beta $
[Eq. (\ref{eq3b})] and $\rho _{1}$ on $d_{t}$. Both Figs. 4 and 5
display rather long (around few seconds) spin relaxation of
localized electrons in CNT at low temperature that makes it
attractive for quantum computing application.

The magnetic field and temperature dependencies of spin-relaxation
rate calculated with Eq.~(\ref{eq12}) is shown in Fig.~6 for the
delocalized electron (i.e., without taking into consideration the
electron scattering effects). Some decreasing attributes to only low
temperature relaxation in actual range of the magnetic fields while
no significant dependence on ${\mathrm B}$ reveals the $T_{1}^{-1}$
starting with $T=40$ K. For the room temperature, Eq.~(\ref{eq12})
predicts the relaxation time about $T_{1}\simeq 30$ $\mu $s in the
wide range of the magnetic fields. The $T_{1}^{-1}$ tends to
increase as the temperature goes up while mechanism due to hyperfine
interaction possesses the opposite tendency. The increase of the
tube diameter suppresses spin relaxation in the case of both
mechanisms (Fig. 7). Note that with the exception of very low
magnetic field, which corresponds to Eq.~(\ref{eq14}), the spin
relaxation via flexural modes dominates over the processes of
electron scattering at nuclei spins. This deduction is based on the
comparison of present results and theory developed in
Ref.~\onlinecite{SemenovHFI2007}, where the constant of HFI has been
specified, $a_{hf}/2\pi \hbar =-35.8$ MHz.~\cite{Yazyev}

\section{Conclusions}

A mechanism of electron spin relaxation in carbon nanotubes caused
by anisotropy of g-tensor and flexural phonon modes is studied.
Relaxation time for localized electron is estimated around a
second at low temperature while the spin relaxation of delocalized
electrons can reach few tens microseconds at room temperature. As
it turned out the proposed mechanism is essentially more efficient
than the processes of spin scattering due to hyperfine interaction
with isotopes $^{13}$C. This holds true up to very weak magnetic
fields when Zeeman splitting comes up with electron energy
broadening. The mechanism reveals specific dependencies on
magnetic field orientation (common for both localized and
delocalized electrons) and on the magnetic field strength (with
perceptible maximum for localized electron and almost flat
dependence for delocalized one). Besides, in contrast to processes
of scattering on nuclei spins, the temperature augmentation
sufficiently increases relaxations rate. These particularities
will facilitate the experimental recognition of the mechanism of
spin-lattice relaxation in CNT.

\begin{acknowledgments}
This work was supported in part by the SRC/MARCO Center on FENA and the
National Science Foundation.
\end{acknowledgments}

\newpage

\begin{center}
\begin{figure}[tbp]
\includegraphics[scale=.7,angle=0]{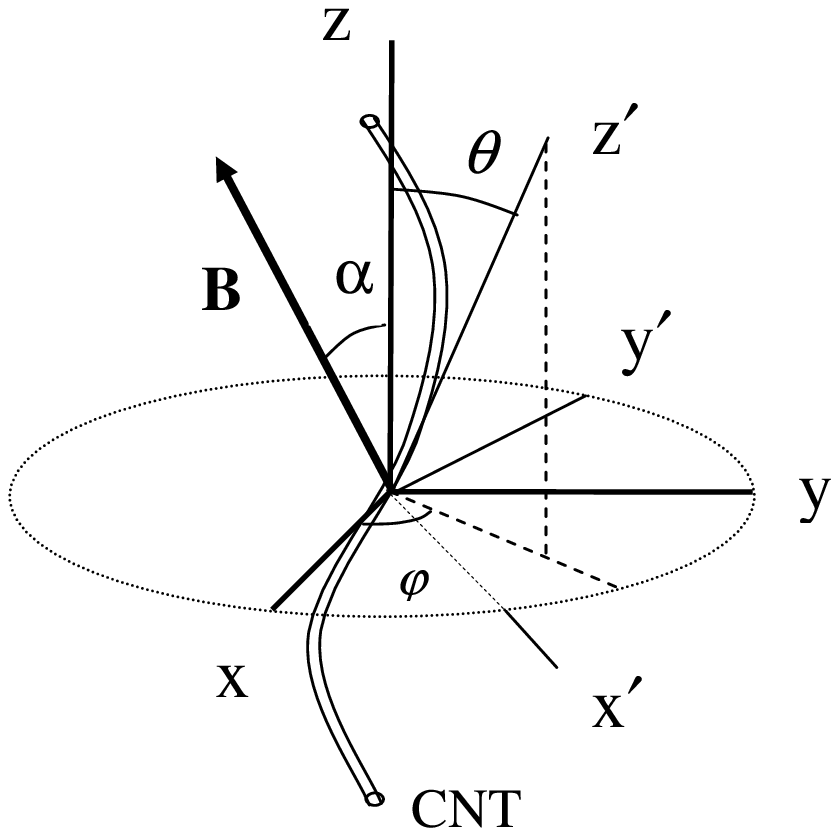}
\caption{The position of the laboratory coordinate system $x$, $y$
and $z$, the local coordinate $x'$, $y'$ and $z'$ related to CNT
flexure and the angles between axes. The direction of a magnetic
field $\textbf{B}$ determines the angle $\alpha$. }
\end{figure}
\end{center}

\newpage

\begin{center}
\begin{figure}[tbp]
\includegraphics[scale=.7,angle=0]{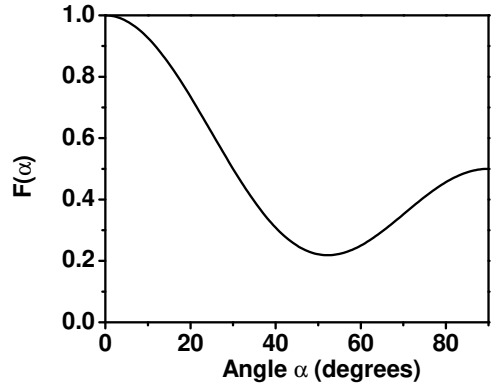}
\caption{Function graph of $F(\alpha)$, which defines the
dependence of spin relaxation rate on the angle $\alpha$ between
the direction of magnetic field and the axis of CNT for both
localized and delocalized electrons.}
\end{figure}
\end{center}

\newpage

\begin{center}
\begin{figure}[tbp]
\includegraphics[scale=.7,angle=0]{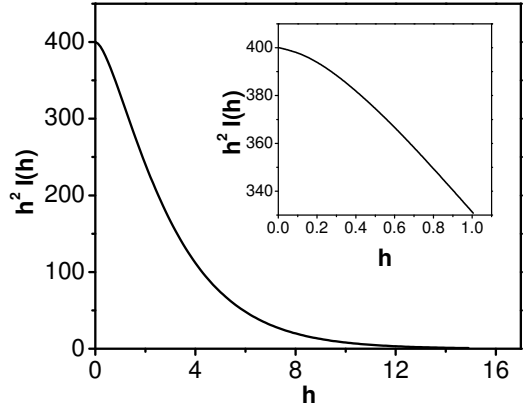}
\caption{Plot of function $h^{2}I(h)$, which appears in Eq.
(\ref{eq12}).}
\end{figure}
\end{center}

\newpage

\begin{center}
\begin{figure}[tbp]
\includegraphics[scale=.65,angle=0]{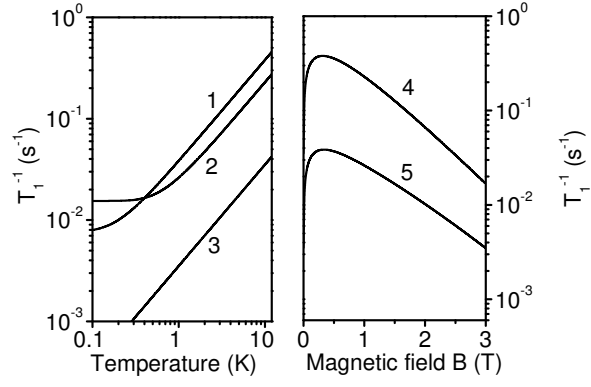}
\caption{Calculated spin relaxation rates of localized electron in
a (8,4) CNT as a function of magnetic field and temperature.
Curves 1, 2, and 3 are obtained for $B=1T,$ $0.3T$, and $0.001T$
correspondingly. Curves 4 and 5 are plotted for $T=10$K and $1$K.
The size of electron localization is $2d_{0} = 30$~nm.}
\end{figure}
\end{center}

\newpage

\begin{center}
\begin{figure}[tbp]
\includegraphics[scale=.65,angle=0]{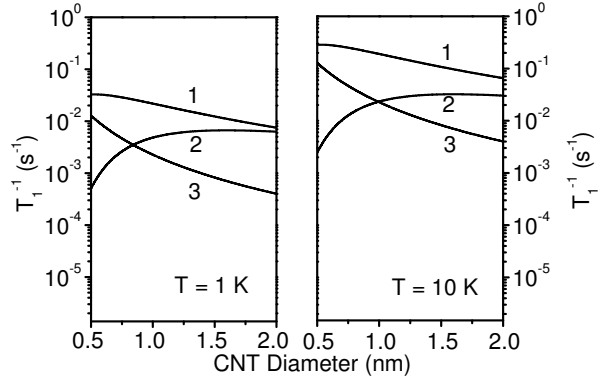}
\caption{Calculated spin relaxation rates of a localized electron
as a function of CNT diameter for different values of magnetic
field and temperature and temperatures $T=1$K (left panel)and
$T=10$ K (right panel). Curves 1, 2, and 3 correspond to
$B=0.001T$, $1T$, and $3T$ respectively. The size of electron
localization is $2d_{0} = 30$ nm.}
\end{figure}
\end{center}

\newpage

\begin{center}
\begin{figure}[tbp]
\includegraphics[scale=.65,angle=0]{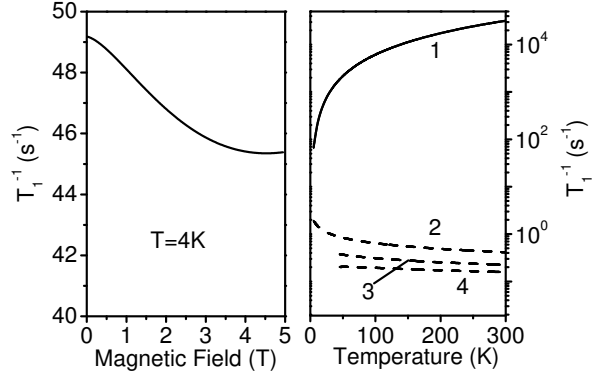}
\caption{Calculated spin relaxation rates for mechanism caused by
flexural modes (solid lines) and due to electron scattering on
$^{13}$C isotopes (dashed line) for delocalized electron in a
(8,4) CNT as a function of magnetic field at $T=4$K (left panel)
and temperature (right panel). The curves 2, 3 and 4 are obtained
for $B=0.05$T, $1$T, and $5$T correspondingly. Solid line 1
practically does not change in range of the magnetic field from
$B=0.05$T up to $B=5$T.}
\end{figure}
\end{center}

\newpage

\begin{center}
\begin{figure}[tbp]
\includegraphics[scale=.65,angle=0]{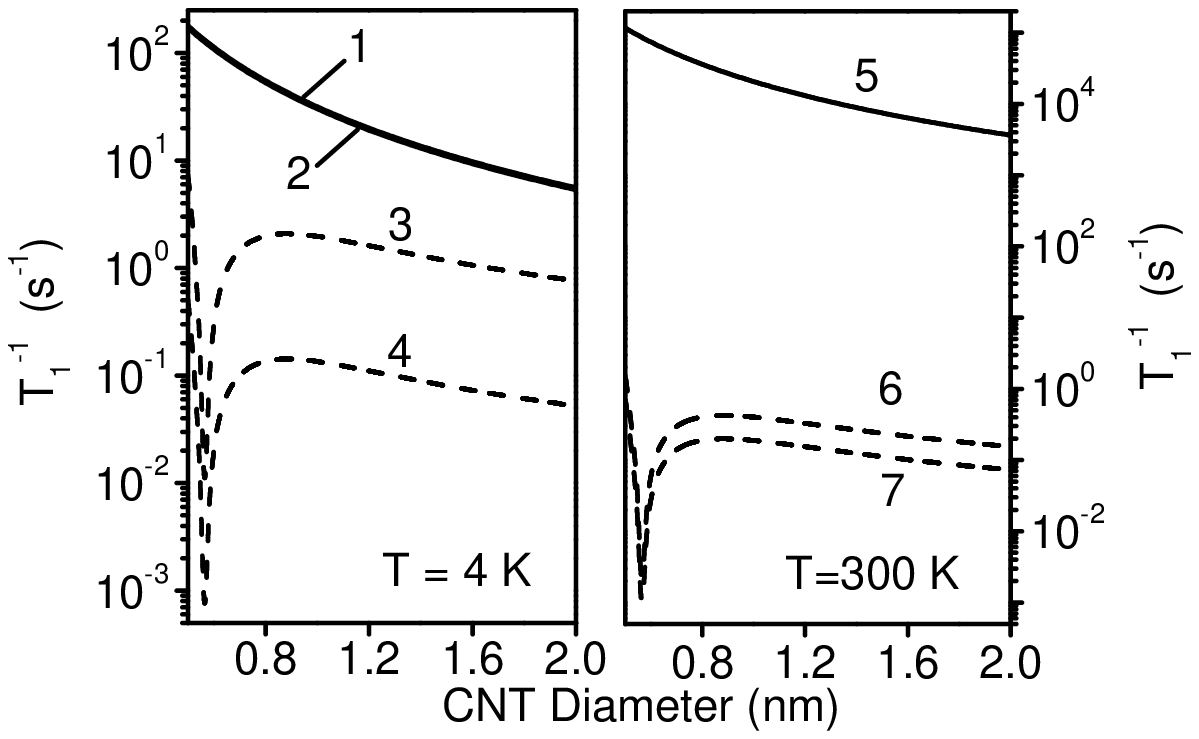}
\caption{Calculated spin relaxation rates of delocalized electron
as a function of CNT diameter for different values of magnetic
field and temperature. Contributions of both HFI (dashed line) and
spin-phonon interaction (solid line) are presented for the purpose
of comparison. Curves 1, 3, and 6 correspond to $B=0.05$T. Curves
2, 4, and 7 correspond to $B=2$T. Curve 5 does not change in the
range of the magnetic field $B=0.05 - 2$T.}
\end{figure}
\end{center}


\newpage
\begin{thebibliography}{99}
\bibitem{Dresselhaus} \textit{Carbon Nanotubes: Synthesis, Structure,
Properties and Applications}, ed. M. S. Dresselhaus, G. Dresselhaus, and P.
Avouris (Springer, Berlin, 2000); M. S. Dresselhaus, G. Dresselhaus, and P.
S. Eklund, \textit{Science of Fullerences and Carnon Nanotubs} (Academic,
New York. 1996).

\bibitem{Ando05} T. Ando, J. Phys. Soc. Japan \textbf{74}, 777 (2005).

\bibitem{Dekker} See, for example, C. Dekker, Phys. Today \textbf{52}, 22
(1999); T. W. Odom, J. L. Huang, P. Kim, and C. M. Lieber, J. Phys. Chem. B
\textbf{104}, 2794 (2000).

\bibitem{Tsukagoshi99} K. Tsukagoshi, B. W. Alphenaar, and H. Ago,
Nature (London) \textbf{401}, 572 (1999).

\bibitem{Sahoo05} S. Sahoo, T. Kontos, and C. Sch\"{o}nenberger, Appl. Phys.
Lett. \textbf{86}, 112109 (2005).

\bibitem{Meh} H. Mehrez, J. Taylor, H. Guo, J. Wang, and C. Roland, Phys.
Rev. Lett. \textbf{84}, 2682 (2000).

\bibitem{Yang} C.-K. Yang, J. Zhao, and J. P. Lu, Phys. Rev. Lett. \textbf{90%
}, 257203 (2003).

\bibitem{LossDiVincenzo} D. Loss and D. P. DiVincenzo, Phys. Rev. A \textbf{%
57}, 120 (1998).

\bibitem{KaneNature1998} B. E. Kane, Nature (London) \textbf{393}, 133-137 (1998).



\bibitem{SemenovHFI2007} Y. G. Semenov, K. W. Kim, G. J. Iafrate, Phys. Rev.
B \textbf{75}, 045429 (2007).

\bibitem{BulaevLoss2005} D. V. Bulaev and D. Loss, Phys. Rev. B
\textbf{71}, 205324 (2005).

\bibitem{Mahan2002} G. D. Mahan, Phys. Rev. B \textbf{65}, 235402 (2002).

\bibitem{Oshima88} C. Oshima, T. Aizawa, R. Souda, Y. Ishizawa, and Y.
Sumiyoshi, Solid State Commun. \textbf{65}, 1601 (1988).

\bibitem{Kittel} C. Kittel, \textit{Quantum Theory of Solids} (Wiley, New York,
1963).

\bibitem{SK07} Y. G. Semenov and K. W. Kim, Phys. Rev. B \textbf{75}, 195342
(2007).

\bibitem{Sapmaz06} S. Sapmaz, P. J.-Herrero, L. P. Kouvenhoven, and H. S. J. van
der Zant, Semicond. Sci. Technol., \textbf{21}, S52 (2006).

\bibitem{Argyres} P. N. Argyres, J. Phys. Chem. Solids \textbf{4}, 19 (1958).

\bibitem{g-factor} O. Chauvet, L. Forro, W. Bacsa, D. Ugarte, B. Doudin, and
W. A. de Heer, Phys. Rev. B \textbf{52}, R6963 (1995).

\bibitem{Yazyev} O. V. Yazyev, private communication. See also O. V.
Yazyev and L. Helm, Phys. Rev. B \textbf{72}, 245416 (2005).
\end{thebibliography}
\end{document}